\documentclass[fleqn,10pt]{wlscirep}
\usepackage[utf8]{inputenc}
\usepackage[T1]{fontenc}
\usepackage{subfigure}
\usepackage{amsmath, amssymb, amsthm}
\usepackage{algorithm}
\usepackage{algpseudocode}
\usepackage{graphicx}
\usepackage{hyperref}
\usepackage{enumitem}
\usepackage{geometry}
\usepackage{float}          
\usepackage{cite}           
\usepackage{bm}             
\usepackage{float}
\usepackage[figureposition=bottom,tableposition=bottom]{caption}
\usepackage{setspace}    
\usepackage{caption}     

\usepackage{etoolbox}
\AtBeginEnvironment{equation}{%
  \setlength\abovedisplayskip{4pt}%
  \setlength\belowdisplayskip{4pt}%
  \setlength\abovedisplayshortskip{4pt}%
  \setlength\belowdisplayshortskip{4pt}%
}

\usepackage{titlesec}
\titlespacing*{\subsection}{0pt}{4pt}{4pt}

\newcommand{\ket}[1]{\left|#1\right\rangle}

\title{Hybrid Quantum–Classical Surrogate for Real-Time Inverse Finite Element Modeling in Digital Twins}

\author[1*,+]{Azadeh Alavi}
\author[2,+]{Sanduni Jayasinghe}
\author[1]{Mojtaba Mahmoodian}
\author[3]{Sam Mazaheri}
\author[2]{John Thangarajah}
\author[1]{Sujeeva Setunge}

\affil[1]{School of Engineering, RMIT University, Melbourne, VIC 3000, Australia}
\affil[2]{School of Computing Technologies, RMIT University, Melbourne, VIC 3000, Australia}
\affil[3]{Dalrymple Bay bulk material Terminal, Hay Point, QLD 4740 and Beta International Associate Pty Ltd.}

\affil[*]{azadeh.alavi@rmit.edu.au}

\affil[+]{these authors contributed equally to this work}

\begin{abstract}
    
Large-scale civil structures, such as bridges, pipelines, and offshore platforms, are vital to modern infrastructure, where unexpected failures can cause significant economic and safety repercussions. Although finite element (FE) modeling is widely used for real-time structural health monitoring (SHM), its high computational cost and the complexity of inverse FE analysis, where low-dimensional sensor data must map onto high-dimensional displacement or stress fields—pose ongoing challenges. Here, we propose a hybrid quantum–classical multilayer perceptron (QMLP) framework to tackle these issues and facilitate swift updates to digital twins across a range of structural applications.

Our approach embeds sensor data using symmetric positive definite (SPD) matrices and polynomial features, yielding a representation well-suited to quantum processing. A parameterized quantum circuit (PQC) transforms these features, and the resultant quantum outputs feed into a classical neural network for final inference. By fusing quantum capabilities with classical modeling, the QMLP handles large-scale inverse FE mapping while preserving computational viability.

Through extensive experiments on a bridge, we demonstrate that the QMLP achieves a mean squared error (MSE) of $3.16\times10^{-11}$, outperforming purely classical baselines with a large margin. These findings confirm the potential of quantum-enhanced methods for real-time SHM, establishing a pathway toward more efficient, scalable digital twins that can robustly monitor and diagnose structural integrity in near-real time.  
\end{abstract}

\begin{document}
\flushbottom
\maketitle
\thispagestyle{empty}

\section*{Introduction}

\textbf{Critical civil structures} (e.g., bridges, long-span pipelines, and offshore platforms) form essential components of modern infrastructure, where even minor deterioration can disrupt vital services and incur high economic costs. While we focus here on a jetty conveyor bridge bulk material terminal, a complex, high-impact structure exposed to harsh environmental loads, the same methodology can be extended to other large-scale assets requiring inverse finite element (FE) modeling in a digital twin context. Indeed, bridges are only one example among many industrial and civil systems that rely on real-time structural health monitoring (SHM).

Hence, SHM is essential, especially in digital twin frameworks that synchronize physical structures with virtual models \cite{su14148664,VANDINTER2022107008,su132011549,inproceedings}. Although digital twins, integrated with FE modeling, provide sophisticated simulations for SHM, they face two main challenges in real-time applications: (1) performing inverse analysis using limited sensor data with high-dimensional outputs \cite{doi:10.1142/S0219876219500452,Gupta2013InverseMF}, and (2) the computational intensity of FEM, which limits continuous monitoring \cite{s25010059}.

Machine learning (ML), particularly artificial neural networks (ANNs), helps address these issues by accelerating FEM-based analyses and facilitating real-time, high-dimensional inverse modeling \cite{GUAN2023105120,civileng6010002,article,JAYASINGHE2024119187}. Three main approaches integrate ANNs with FEM: embedding ANNs into FEM (e.g., mesh generation \cite{https://doi.org/10.1002/nme.6235}, material modeling \cite{TAO2022114548}); using ANNs to derive empirical equations from FEM data \cite{TOHIDI201648}; and having FEM generate training data for ANNs to learn direct input-output mapping \cite{SHAHANI20091920}. Although physics-informed neural networks (PINNs) bolster physical interpretability, they often demand high computational resources \cite{MITUSCH2021110651,article}. Our strategy adopts the latter FEM-to-ANN approach but uniquely combines quantum computing for enhanced efficiency and scaling.

Common ANN architectures for this integration include multi-layer perceptrons (MLPs) \cite{app11209411,GUAN2023105120,LESHNO1993861}, recurrent neural networks (RNNs) \cite{CAO2020102869,GREVE2022100137}, convolutional neural networks (CNNs) \cite{articlecnn,articlecnn2}, and graph neural networks (GNNs) \cite{a16080380,Taghizadeh2024MultifidelityGN}. While MLPs are widely used for their simplicity, large-scale FEM tasks involving big meshes often strain classical computational resources \cite{IBRAGIMOVA2022103374,GREVE2022100137}. More advanced architectures like GNNs can better represent mesh-based data, yet conventional computing methods may still limit their performance \cite{Taghizadeh2024MultifidelityGN}.

Quantum computing (QC) exploits quantum mechanics principles to achieve computational tasks exceeding classical capabilities. In civil engineering, QC can accelerate simulations, optimizations, and ML methods for large-scale structural analysis \cite{ploennigs2024quantumcomputingcivilengineering,LIU2024102117,articleSato,Montanaro_2016}. Quantum machine learning (QML) merges quantum hardware advantages with classical ML, offering a promising hybrid approach for real-time inverse FEM in SHM \cite{Biamonte2016QuantumML,Alavi2024LeveragingSPD}. Although QML remains in early development, its potential for enhancing inverse FEM and digital twin applications is vast, marking a research gap this study addresses.

In this work, a novel hybrid quantum-classical MLP (QMLP) pipeline is proposed to tackle the real-time inverse FE problem. The method leverages symmetric positive definite (SPD) matrices on Riemannian manifolds, aligning classical data more naturally with quantum states while preserving essential structural properties. Polynomial feature expansion is incorporated to capture complex nonlinear relationships.\\

\textbf{Contributions:} The core contributions of this work are threefold.
\begin{itemize}
    \item We introduce a hybrid quantum-classical neural network architecture for real-time inverse FE modeling, addressing the dimensional gap between low-dimensional sensor data and high-dimensional structural responses.
    \item We develop an SPD-based feature embedding on polynomial expansions, which enhances data conditioning and enables an efficient parameterized quantum circuit (PQC) to learn nonlinear relationships essential for accurate displacement predictions. We also integrate a Hilbert–Schmidt embedding into the quantum feature pipeline to capture complex system dynamics in a high-dimensional space, thereby enhancing synergy between quantum and classical components.
    \item We validate our method on bridge data, demonstrating superior performance (MSE $3.1\times10^{-4}$) compared to purely classical baselines, and discuss its implications for scalable digital twin implementations.
\end{itemize}

\vspace{2mm}
\noindent
These contributions strengthen the case for integrating quantum processing layers into classical SHM pipelines, paving the way for high-fidelity, real-time simulation and anomaly detection in critical structures.

\section*{Mathematical Background}
This section outlines the key mathematical concepts that support our proposed quantum machine learning approach for real-time FE modeling. We first discuss the instability inherent in conventional inverse modeling, then present the role of Symmetric Positive Definite (SPD) matrices and their Riemannian geometry, and finally review essential aspects of quantum encoding relevant to inverse FE analysis.

\subsection*{(i) Inverse Modeling and SPD Matrices}
Inverse modeling in FE analysis refers to estimating internal structural responses (e.g., stresses, strains, displacements) from limited observed or measured data. Unlike forward modeling, which applies known loads/boundary conditions to predict structural behavior, inverse modeling reconstructs the entire response without direct knowledge of the applied loads. Let $\mathbf{y}\in \mathbb{R}^{m}$ denote sensor measurements, $\mathbf{x}\in \mathbb{R}^{n}$ the unknown structural responses, and $\mathbf{A}$ a transformation matrix, with $\epsilon$ representing measurement noise; thus:  $\mathbf{y} = \mathbf{A}\mathbf{x} + \epsilon.$

Since instrumentation of an entire structure is typically expensive and impractical, $\mathbf{y}$ is underdetermined with respect to $\mathbf{x}$. This null-space ambiguity gives rise to an infinite set of solutions unless additional constraints are imposed. Mathematically:
\vspace{0.5mm}
\begin{equation}
\mathbf{A}^T \mathbf{A} \mathbf{x} = \mathbf{A}^T \mathbf{y}, \qquad
\mathbf{x} = \mathbf{x}_p + \mathbf{x}_n,
\qquad
\frac{\|\delta \mathbf{x}\|}{\|\mathbf{x}\|} \leq \kappa(\mathbf{A}^T \mathbf{A}) \frac{\|\delta \mathbf{y}\|}{\|\mathbf{y}\|}.
\end{equation}
\vspace{0.5mm}
where $\mathbf{x}_p$ is a particular solution and $\mathbf{x}_n \in \mathcal{N}(\mathbf{A})$ is any null-space vector. A rank-deficient $\mathbf{A}$ often leads to high condition numbers, making the system sensitive to small perturbations in $\mathbf{y}$.

Hence, slight sensor noise can yield large errors in the inverse solution, complicating direct inversion. Neural network surrogates trained on simulated data provide a more robust alternative: they learn an approximate mapping $\mathbf{y}\mapsto \mathbf{x}$ without explicit matrix inversion. 

To stabilize this learning process, we utilize \emph{Symmetric Positive Definite} (SPD) matrices. An SPD matrix $\mathbf{S}$ is symmetric with strictly positive eigenvalues. Formally, $\mathbf{P} = \mathbf{P}^\top$ and $\mathbf{y}^\top\mathbf{P}\,\mathbf{y} > 0$ for all nonzero $\mathbf{y}$. The set of all SPD matrices $\mathcal{S}_{++}^n$ forms a Riemannian manifold, allowing us to exploit specialized geometric tools (e.g., Riemannian metrics) to manage high-dimensional transformations with greater numerical stability. As detailed below, we embed sensor data into SPD matrices to facilitate robust feature representations and quantum encoding.

\subsection*{(ii) Spectral Decomposition and Feature Transformation}
Feature transformation improves expressiveness while controlling computational overhead. We employ polynomial feature expansion plus dimensionality reduction via eigenvalue decomposition, using the geometric and spectral properties of SPD matrices. 

First, a polynomial expansion maps the $d$-dimensional input $\mathbf{x}\in \mathbb{R}^d$ to a higher-dimensional space by including all monomials up to a chosen degree $p$. For instance, a second-degree expansion ($p=2$) yields:
\begin{equation}
\mathbf{z} = [\,y_1,\;y_2,\dots,y_d,\;y_1^2,\;y_1y_2,\dots,y_d^2\,]^\top \in \mathbb{R}^{d'}.
\end{equation}
This step captures crucial nonlinear interactions in structural data. However, when $d'$ is large, a dimensionality reduction can be beneficial. Given an SPD matrix $\mathbf{P}\in \mathbb{R}^{d'\times d'}$, we factor it via:
\vspace{0.5mm}
\begin{equation}
\mathbf{P} = \mathbf{V}\,\Lambda\,\mathbf{V}^\top,
\end{equation}
\vspace{0.5mm}
where $\mathbf{V}$ is orthogonal and $\Lambda$ holds strictly positive eigenvalues. Retaining the top $k$ eigenvalues $\lambda_i$ yields a reduced representation:
\begin{equation}
\mathbf{y}_{\text{proj}} = \mathbf{V}_k^\top\,\mathbf{z},
\end{equation}
where $\mathbf{V}_k\in \mathbb{R}^{d'\times k}$ contains the leading eigenvectors, effectively compressing data while preserving key structural information.

\subsection*{(iii) Quantum State Preparation and Encoding}
Quantum state preparation converts classical data into quantum states. We adopt \emph{amplitude encoding}, which maps a normalized vector $\mathbf{y}\in \mathbb{R}^n$ (with $\|\mathbf{y}\|=1$) to a quantum state:
\begin{equation}
\ket{\psi_y} = \sum_{i=1}^{n} y_i \ket{i}.
\end{equation}
This superposition allows quantum processors to handle high-dimensional data more efficiently than classical methods. In our approach, we first embed classical vectors into SPD representations on a Riemannian manifold, then generate a structured quantum encoding. By aligning with the geometric framework of quantum mechanics, we preserve essential features and enable more effective quantum state transitions, ultimately improving computational efficiency for inverse FE analysis in a hybrid quantum–classical machine learning setting.

\section{Case Study and Data Preparation}
\label{sec:case_study}

In order to assess the effectiveness of the proposed hybrid quantum-classical multilayer perceptron (QMLP), we conducted a case study on a bridge. This section describes the practical motivation for selecting this structure, the data acquisition process, and the FE-based dataset generation that underpins our QMLP training. The subsequent methodology (Section~\ref{sec:methodology}) then details how these data are transformed and fed into our hybrid model.

\subsection{Bridge Description and Sensor Configuration}
\label{subsec:bridge_description}
\begin{figure}[ht]
    \centering
    \includegraphics[width=0.65\linewidth]{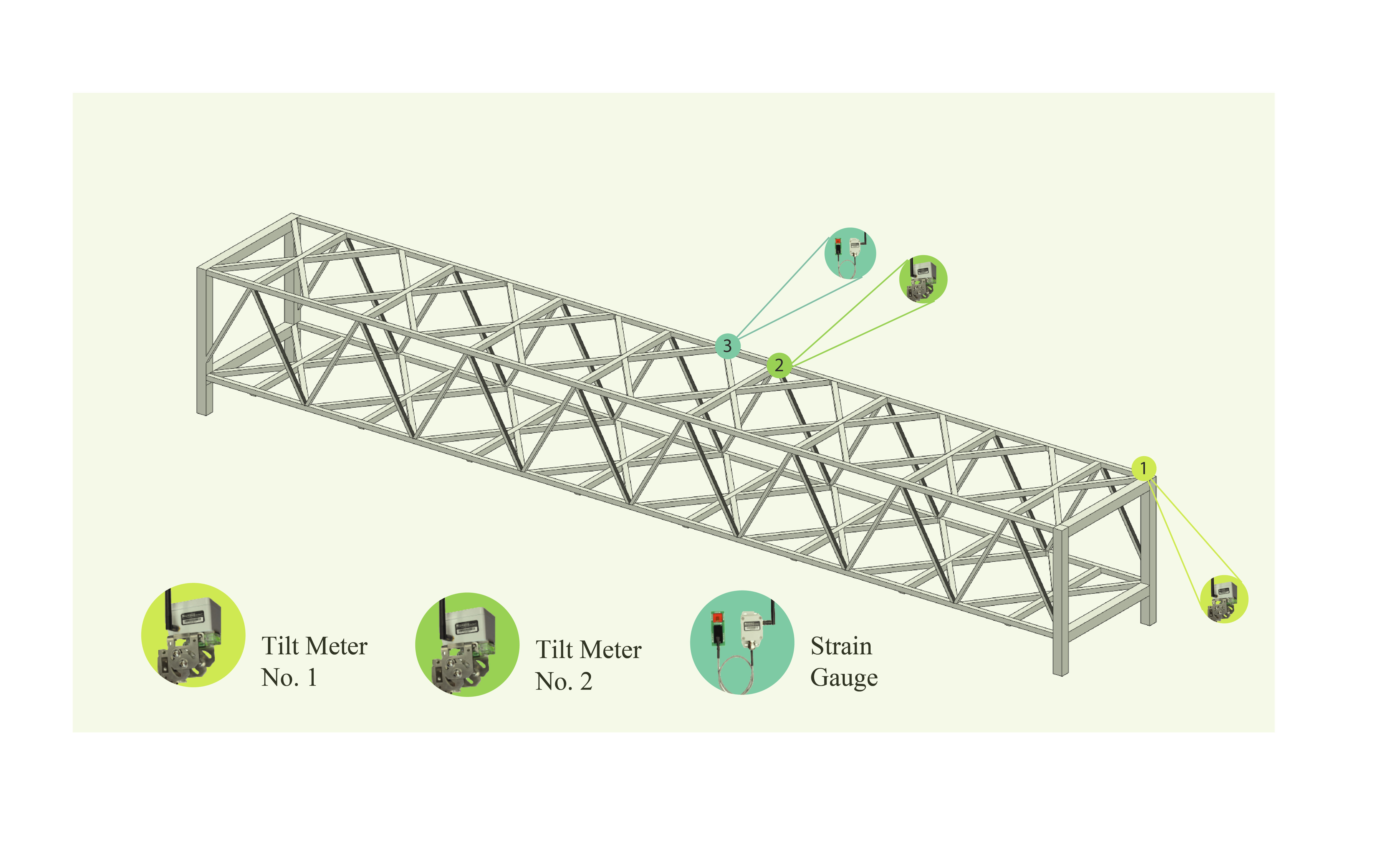}
    \caption{Schematic of the jetty conveyor bridge with sensor positions (S1, S2, and S3).}
    \label{fig:SensorLocations}
\end{figure}

A simplified schematic of the investigated section of the bridge is shown in Fig.~\ref{fig:SensorLocations}. The steel frame under study spans 23.65\,m in length, 3.6\,m in width, and 2.6\,m in height, with hollow-section members of varying cross-sectional dimensions. Three tilt meters (labeled S1, S2, and S3) are deployed at locations deemed most critical based on prior finite element (FE) simulations and structural assessments, each measuring tilt along the $x$, $y$, and $z$ axes. Altogether, these sensors produce 7 input variables (since one axis was deemed redundant or combined), recorded at 15-second intervals. All sensor data are uploaded via a cloud platform for real-time monitoring. 

\subsection{FE Modeling and Dataset Generation}
\label{subsec:fe_model}
To link the sensor measurements to the structure's internal state, we built and validated an FE model of the conveyor segment using Ansys\footnote{Ansys~\textregistered, version 2022R2.}, specifically the static structural design module. The model included:
\begin{itemize}[leftmargin=*]
    \item \textbf{Geometry and Boundary Conditions:} The jetty conveyor section was idealized with shell or beam elements (depending on cross-section), and fixed supports at the four corners (consistent with on-site constraints).
    \item \textbf{Loading Scenarios:} bulk material was assumed to travel along an overhead rail at speeds between 1.5\,m/s and 8.5\,m/s, producing distributed loads between 1.2\,kN/m and 6.5\,kN/m on the top frame. Wind loads ranged from 0 to 54\,m/s, with speeds above 19\,m/s implying no bulk material transport (to align with operational safety limits). {Although these loads are dynamic in nature, in order to mimic a range of loading instances to train the MLP model, they were considered as a series of static loads.\cite{JAYASINGHE2024119187}}
    \item \textbf{Tidal Effects:} Tidal forces were largely negligible in normal operation since the conveyor sits above the high-water mark, although brackish conditions still contribute to corrosion over time.
\end{itemize}

By applying thousands of random load combinations within these physical limits, we generated $10{,}000$ distinct loading cases for the FE model. For each scenario, we captured \emph{both} the resulting nodal displacements (in 3 directions per node) and the simulated sensor readings at the modeled tilt meter locations. This yielded a high-fidelity input-output mapping:
\[
\underbrace{\mathbf{y} \in \mathbb{R}^7}_\text{7 sensor measurements}
\quad\longmapsto\quad
\underbrace{\mathbf{x} \in \mathbb{R}^{1017}}_\text{339 nodes $\times$ 3 directions}.
\]
Hence, our training dataset consisted of $10{,}000$ pairs $(\mathbf{y}^{(i)}, \mathbf{x}^{(i)})$, each reflecting a unique loading configuration.

\subsubsection{FE Model Validation}
\label{subsubsec:fe_validation}
Prior to dataset generation, the FE model was validated against limited real-world sensor readings during several controlled load scenarios (e.g., known bulk material load, measured wind). As detailed in~\cite{JAYASINGHE2024119187}, the model predictions of displacement closely matched the measured responses, confirming that the FE-based approach reliably captures structural behavior within the operational loading envelope.

\subsection{Initial MLP Architecture and Rationale}
\label{subsec:initial_mlp}

Before introducing quantum features, we tested a conventional multi-layer perceptron (MLP) on this dataset to establish a baseline. The MLP had:
\begin{itemize}[leftmargin=*]
    \item \textbf{Input Layer:} 7 input neurons corresponding to the 7 sensor readings.
    \item \textbf{Hidden Layers:} two hidden layers with 64 and 32 neurons, respectively, each with ReLU activation.
    \item \textbf{Output Layer:} 1017 neurons to predict 3D displacements for all 339 nodes.
\end{itemize}
Although this initial MLP provided a rudimentary inverse FE surrogate, it struggled with the structure's high-dimensional output space. Nonlinearities related to wind-bulk material interaction and potential partial correlation among sensor channels hinted that a more expressive model could yield better accuracy.

\subsection{Motivation for Hybrid Quantum-Classical Modeling}
\label{subsec:motivation}
Given the complexity and scale of the jetty conveyor dataset, our interest turned to quantum-enhanced architectures. The notion is that polynomial-based feature expansions, combined with symmetric positive definite (SPD) embeddings and a parameterized quantum circuit, may more effectively capture intricate nonlinearities than purely classical networks. The following section (Section~\ref{sec:methodology}) outlines the details of our SPD-based quantum feature pipeline and the training of the hybrid QMLP. 

\section{Methodology}
\label{sec:methodology}

We propose a hybrid quantum-classical approach to solve the inverse finite element problem in real-time, enabling continuous updating of a digital twin with incoming sensor data. {While the finite element (FE) simulation model under linear static assumptions is utilized for simulation, the inverse problem that predicts full-field displacements from sparse sensor inputs is highly nonlinear in character due to redundancy and correlation across channels (e.g., two or more directions of tilt), environmental coupling such as wind-induced displacement coupling with operational loads (e.g., sliding coal) and the intrinsic ill-posedness of inverse FE analysis with limited measurement points \cite{s25113513, JAYASINGHE2025107698}.} The methodology consists of several key steps: first, we map the measured response data into an enriched feature space and construct a symmetric positive-definite (SPD) matrix representation; second, we convert this SPD matrix into a quantum-compatible density matrix and embed it as a quantum state; third, we design a variational quantum circuit to process this state and extract informative features via quantum measurements; fourth, we integrate these quantum features into a classical neural network that produces the final estimates of the model parameters; and finally, we deploy the trained hybrid model for real-time inference in the digital twin. In the following, we detail each step of the methodology with mathematical precision and clarity.

\subsection{Nonlinear Feature Expansion and SPD Matrix Construction}

Let $\mathbf{x}\in\mathbb{R}^m$ denote the vector of $m$ measured response features from the physical system (e.g., displacements or strains at sensor locations). To capture nonlinear correlations in these inputs, we first transform $\mathbf{x}$ via a polynomial feature expansion. We define $\Phi(\mathbf{x})\in\mathbb{R}^d$ as the expanded feature vector containing all monomials of the components of $\mathbf{x}$ up to a chosen degree (including cross-terms). For example, if $m=2$ and we include up to quadratic terms, $\Phi(x)$ would include $x_1$, $x_2$, $x_1^2$, $x_1 x_2$, $x_2^2$, and a constant bias term. This nonlinear feature mapping $\mathbf{x}\mapsto \Phi(\mathbf{x})$ allows the model to represent complex relationships that a linear model could not capture, which is crucial for the inverse modeling of nonlinear systems.

Next, we use the expanded features to construct a symmetric positive-definite matrix that will later serve as a quantum-compatible encoding. Specifically, we form the outer product (Gram matrix) of the feature vector with itself:
\begin{equation}
    K(\mathbf{x}) \;=\; \Phi(\mathbf{x})\,\Phi(\mathbf{x})^T\,,
    \label{eq:gram}
\end{equation}
where $K(\mathbf{x})\in\mathbb{R}^{d\times d}$. By construction, $K(\mathbf{x})$ is symmetric and positive semidefinite (all its eigenvalues are non-negative). In order to ensure strict positive definiteness (and to improve numerical stability), we add a small regularization term $\varepsilon$ on the diagonal:
\begin{equation}
    K_{\varepsilon}(\mathbf{x}) \;=\; K(\mathbf{x}) \;+\; \varepsilon I_d\,,
    \label{eq:reg}
\end{equation}
where $I_d$ is the $d\times d$ identity matrix and $\varepsilon>0$ is a small constant (e.g., $10^{-6}$). The matrix $K_{\varepsilon}(\mathbf{x})$ is symmetric positive-definite, meaning all eigenvalues $\lambda_i$ satisfy $\lambda_i>0$. This SPD matrix serves two purposes: 
\begin{enumerate}[label=(\alph*), itemsep=0pt, topsep=0pt, parsep=0pt, partopsep=0pt]
\item It encodes the enriched feature information in a quadratic form, implicitly mapping $\mathbf{x}$ into a higher-dimensional feature space (a common trick in kernel methods to capture nonlinearity).
\item It provides a mathematically valid density operator once normalized, which is essential for quantum state embedding as described next.
\end{enumerate}

To interface with quantum computing frameworks, we interpret $K_{\varepsilon}(\mathbf{x})$ as an unnormalized density matrix and convert it into a proper density matrix. We first take the matrix square root of $K_{\varepsilon}(\mathbf{x})$, denoted $\sqrt{K_{\varepsilon}(\mathbf{x})}$, which is well-defined and also SPD (it has the same eigenvectors as $K_{\varepsilon}$, with eigenvalues $\sqrt{\lambda_i}$). We then normalize this matrix to have unit trace:
\begin{equation}
    \rho(\mathbf{x}) \;=\; \frac{\sqrt{K_{\varepsilon}(\mathbf{x})}}{\mathrm{Tr}\!\Big(\sqrt{K_{\varepsilon}(\mathbf{x})}\Big)}\,,
    \label{eq:density}
\end{equation}
where $\mathrm{Tr}(\cdot)$ denotes the matrix trace. By construction, $\rho(\mathbf{x})$ is Hermitian, positive semidefinite, and has $\mathrm{Tr}(\rho)=1$, hence it is a valid density matrix. The small regularization $\varepsilon$ further ensures that $\rho(\mathbf{x})$ is full-rank and well-conditioned. 

\subsection{Quantum State Embedding via Hilbert-Schmidt Mapping}

Given the density matrix $\rho(\mathbf{x})$ obtained above, the next step is to embed this classical information into a quantum state that can be processed by a quantum circuit. One theoretical approach is to utilize the Hilbert-Schmidt isomorphism between linear operators and vectors. In particular, any $d\times d$ density matrix can be mapped to a pure state in a $d^2$-dimensional Hilbert space by vectorizing the matrix. Let $\{\ket{i}\}_{i=1}^d$ be an orthonormal basis for the $d$-dimensional feature space. The Hilbert-Schmidt (column-wise) vectorization of $\rho(\mathbf{x})$ is defined as:
\begin{equation}
    \ket{\psi(\mathbf{x})} \;=\; \frac{1}{\sqrt{\mathrm{Tr}\!\big(\rho(\mathbf{x})^2\big)}} \sum_{i=1}^{d}\sum_{j=1}^{d} [\rho(\mathbf{x})]_{ij}\; \ket{i}\otimes\ket{j}\,,
    \label{eq:vec}
\end{equation}
where $[\rho(\mathbf{x})]_{ij}$ are the elements of $\rho(\mathbf{x})$ in the basis $\{\ket{i}\}$, and $\ket{i}\otimes\ket{j}$ denotes the tensor product of basis states (which spans a $d^2$-dimensional space). The normalization factor ensures that $\ket{\psi(\mathbf{x})}$ has unit norm. By this construction, $\ket{\psi(\mathbf{x})}$ is a legitimate quantum state that encodes all entries of $\rho(\mathbf{x})$. 

If $d=2^n$ is a power of two, then $\ket{\psi(\mathbf{x})}$ lives in a Hilbert space of dimension $2^{2n}$, which corresponds to $2n$ qubits. In practice, directly preparing the state $\ket{\psi(\mathbf{x})}$ may be resource-intensive. Instead, we adopt a more hardware-efficient approach in which $\rho(\mathbf{x})$ guides how we set up the initial configuration of a quantum circuit, typically via \emph{angle encoding}. The next subsection details this variational quantum circuit architecture.

\begingroup
  \small
  \setlength{\abovedisplayskip}{3pt}
  \setlength{\belowdisplayskip}{3pt}
  \setlength{\jot}{3pt}            
  \setlength{\textfloatsep}{6pt}
  \setlength{\floatsep}{6pt}
  \captionsetup{font=small,skip=3pt}

  \section*{Pseudocode: SPD Mapping from Polynomial Features}

  \begin{algorithm}[H]
    \caption{SPD Mapping from Polynomial Features}
    \label{alg:spd_mapping_poly}
    \begin{algorithmic}[1]
      \Require 
        $x \in \mathbb{R}^7$ \Comment{Raw sensor/feature vector of dimension 7}
        \Statex \quad Polynomial degree $p = 3$
        \Statex \quad Regularization constant $\epsilon > 0$
      \Ensure 
        Quantum state $|\psi\rangle \in \mathbb{R}^{d^2}$
        
      \State \textbf{Step 1: Polynomial Expansion} 
      \Statex \quad Compute $\mathbf{z} = \Phi(x)$ using a degree-$p$ polynomial expansion 
      \Statex \quad (excluding bias). 
      \Statex \quad \emph{Notation:} if $d$ is the dimension of the expanded vector then 
                  $\mathbf{z}\in\mathbb{R}^d$.

      \State \textbf{Step 2: SPD Matrix Construction}
      \Statex \quad Form $M\gets \mathbf{z}\,\mathbf{z}^\top$ 
      \Statex \quad Regularize: 
      \[
        M_{\mathrm{SPD}}\gets M + \epsilon\,I_d
      \]

      \State \textbf{Step 3: Spectral Decomposition \& Square Root}
      \Statex \quad $M_{\mathrm{SPD}} = U\,\Lambda\,U^\top$
      \Statex \quad $M^{\tfrac12} = U\,\Lambda^{\tfrac12}\,U^\top$

      \State \textbf{Step 4: Density Matrix Formation}
      \Statex \quad $\rho \gets M^{\tfrac12}/\mathrm{tr}(M^{\tfrac12})$

      \State \textbf{Step 5: Hilbert–Schmidt Vectorization}
      \Statex \quad $v=\mathrm{vec}(\rho)\in\mathbb{R}^{d^2}$
      \Statex \quad $|\psi\rangle=v/\|v\|$

      \State \Return $|\psi\rangle$
    \end{algorithmic}
  \end{algorithm}
\endgroup

\subsection{Variational Quantum Circuit Architecture}
{Quantum neural networks (QNNs) are variational quantum circuits that mirror classical neural architectures, enabling nonlinear data processing on quantum hardware. Early models date back to Altaisky’s foundational proposal \cite{Altaisky2001Quantum}, and Schuld \emph{et al.} laid out the core requirements and challenges in 2014 \cite{Schuld2014Quest}. Continuous-variable implementations further expanded the model space \cite{Killoran2018Continuous}, though training on NISQ devices can encounter “barren plateaus” in the optimization landscape \cite{McClean2018Barren}. For a thorough survey of methods and circuit designs, see Zhao and Wang \cite{Zhao2021Review}.}
After embedding the data into a quantum state, we employ a parameterized quantum circuit (PQC) to process the information and extract features useful for predicting the unknown parameters of the system. Our quantum circuit architecture is illustrated in Fig.~3. It consists of $n$ qubits and is structured in two parts: an initial data embedding layer and a series of trainable entangling layers.

\paragraph{Data embedding.}
We begin with all qubits initialized in the standard $\ket{0}$ state. We then apply a set of single-qubit rotation gates to encode the classical data. One approach is \textit{angle embedding}, where each selected feature of the input is mapped to a rotation angle $R_y(\theta_i)$ on qubit $i$. This prepares an initial state $|\Phi(\mathbf{x})\rangle$ that reflects the input data. 

\begin{figure}[h]
    \centering
    \includegraphics[width=0.65\textwidth]{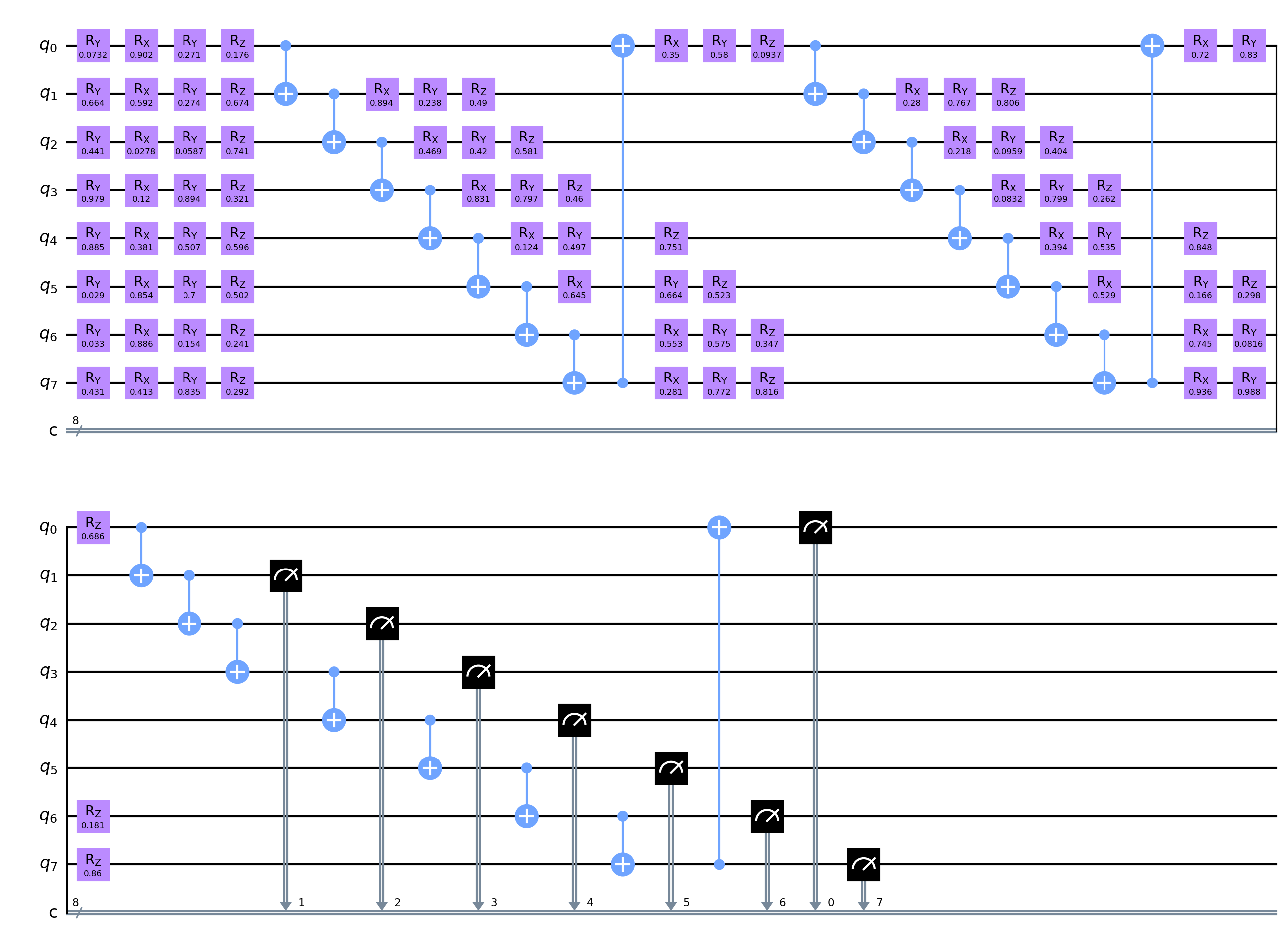}
    \caption{. Structure of the parameterized quantum circuit used in our hybrid model. }
    \label{fig:quantum_circuit_advanced}
\end{figure}

\paragraph{Entangling layers.}
We then apply $L$ layers of a trainable variational ansatz that entangles the qubits and transforms the state in a complex, nonlinear fashion. Each layer $\ell$ applies a sequence of parameterized rotations $R_{\alpha}(\theta_{\ell,i}^{(\alpha)})$ on each qubit $i$, followed by entangling gates (e.g., CNOT) between selected pairs of qubits. The final state of the quantum circuit is 
\[
|\Psi_{\mathrm{out}}(\mathbf{x})\rangle \;=\; U_L(\boldsymbol{\theta}_L)\,\cdots\,U_1(\boldsymbol{\theta}_1)\;|\Phi(\mathbf{x})\rangle.
\]
By adjusting the parameters $\boldsymbol{\theta} = \{\theta_{\ell,i}^{(\alpha)}\}$, the circuit can approximate a wide range of mappings from the embedded state to a desired output state.
Figure. \ref{fig:quantum_circuit_advanced} illustrates the parameterized quantum circuit used in our hybrid model with angle embedding and ring-based entangling layers implemented in Qiskit.

\subsection*{Quantum Measurements with PennyLane}
In our implementation, the \emph{PennyLane} library automates both the quantum circuit simulation and the measurement of expectation values. Specifically, we define a parameterized quantum circuit (PQC) on \texttt{default.qubit} (or another suitable device), and then use PennyLane’s built-in \texttt{qml.expval} measurements to obtain the expectation values of selected Pauli observables. For instance, measuring the Pauli-Z operator on each qubit returns an $n$-dimensional vector 
\[
\langle Z_1 \rangle, \langle Z_2 \rangle, \ldots, \langle Z_n \rangle
\]
which is then passed to the classical neural network layers for final regression. Because PennyLane automatically manages the circuit execution and measurement, users can focus on model design without handling gate-by-gate measurement details.

\subsection{Hybrid Quantum-Classical Network Integration and Training}

The quantum circuit alone does not directly output the desired finite element parameters; instead, it produces a transformed feature vector. We integrate this quantum feature extraction with a classical neural network to learn the mapping from the quantum features to the target parameters (e.g., high-dimensional displacements or other structural variables). The combined system is trained end-to-end via a mean squared error (MSE) loss function:
\begin{equation}
    \mathcal{L} \;=\; \frac{1}{N}\sum_{k=1}^{N}\Big\|\hat{\mathbf{y}}(\mathbf{x}^{(k)}) - \mathbf{y}^{(k)}\Big\|^2,
\end{equation}
where $\hat{\mathbf{y}}(\mathbf{x}^{(k)})$ denotes the model's prediction for the $k$-th sample. Gradients with respect to both classical and quantum circuit parameters are computed via backpropagation and parameter-shift rules, respectively, and we use stochastic gradient-based optimizers such as Adam. We also employ early stopping and $L_2$-regularization to mitigate overfitting.

\paragraph{Clustering-based hyperparameter tuning.}
To determine appropriate sizes for hidden layers, we apply $k$-means clustering on either the quantum output embeddings or the target space. The chosen number of clusters ($k$) informs the number of neurons in the penultimate layer, ensuring that the network capacity aligns with the complexity of the inverse problem.

\subsection{Real-Time Digital Twin Inference Workflow}

Once trained, the hybrid model is deployed to perform real-time inference as part of the digital twin system:
\begin{enumerate}
\item \textbf{Data acquisition:} Acquire sensor data $\mathbf{x}$ from the physical structure.
\item \textbf{Feature expansion:} Compute $\Phi(\mathbf{x})$ and form $K_{\varepsilon}(\mathbf{x})$, then obtain $\rho(\mathbf{x})$ via Eq.~(\ref{eq:density}).
\item \textbf{Quantum processing:} Encode the data into the PQC (e.g., angle embedding), run the circuit, and measure observables to get quantum feature vector $\mathbf{z}$.
\item \textbf{Classical mapping:} The neural network takes $\mathbf{z}$ as input and outputs the predicted FE parameters or responses $\hat{\mathbf{y}}(\mathbf{x})$.
\item \textbf{Update twin:} Use $\hat{\mathbf{y}}(\mathbf{x})$ to update the finite element model in the digital twin, enabling near real-time structural assessment.
\end{enumerate}
This procedure offers significantly reduced latency relative to classical iterative solvers, thus enhancing the timeliness and fidelity of digital twin predictions.

\section*{Results and Discussion}
This section presents a comprehensive evaluation of the proposed hybrid quantum-classical model, highlighting the comparative performance of various architectural choices and providing insights into how each component (from clustering-based dimension analysis to quantum encoding) contributes to improved predictive accuracy. The discussion here is framed in two main parts: (1)~the application of clustering for optimizing classical neural network design, and (2)~the integration of quantum feature encoding schemes, culminating in the proposed \textit{QMLP (Poly-SPD and Hilbert space)} model. All results reference the structural response prediction task described earlier, with emphasis on mean-squared error (MSE), root-mean-squared error (RMSE), and coefficient of determination ($R^2$).

Our results underscore that hybrid QML frameworks, are not mere theoretical constructs but can indeed boost performance for real-world engineering tasks such as bridge FEM. Although we used a quantum simulator here, the approach is designed to transfer to NISQ or future fault-tolerant quantum devices, offering a potentially scalable route to real-time structural health monitoring.

\subsection*{1. Clustering Analysis of Input Data}
\label{subsec:clustering}
\begin{figure}[ht]
    \centering
    \subfigure[Optimum cluster selection by elbow method]{
        \includegraphics[width=0.4\textwidth]{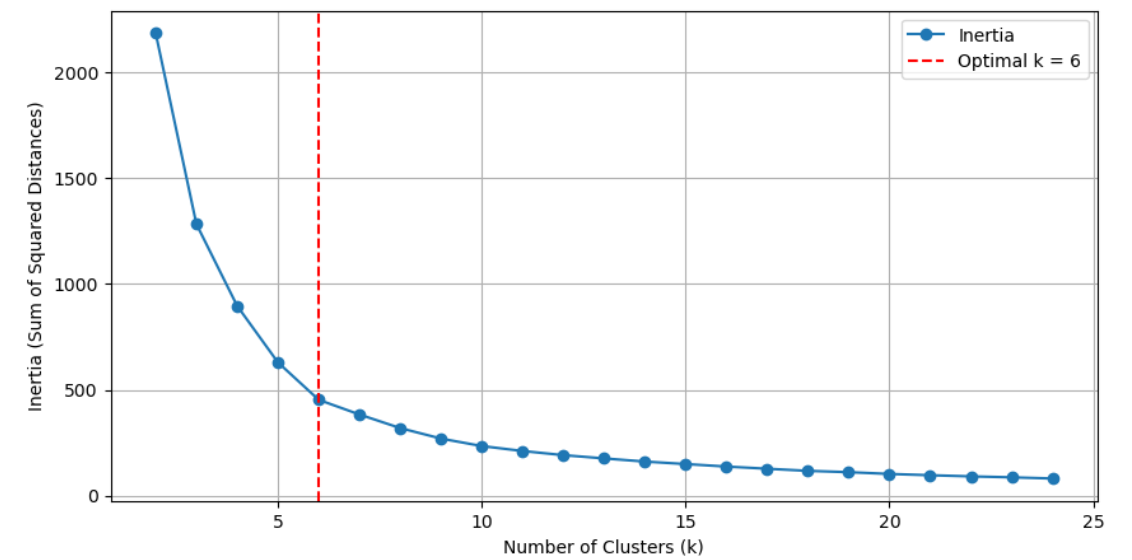}
        \label{fig:subfig1}
    }
    \hspace{0.02\textwidth}
    \subfigure[Optimal Cluster Selection Using Silhouette Score]{
        \includegraphics[width=0.4\textwidth]{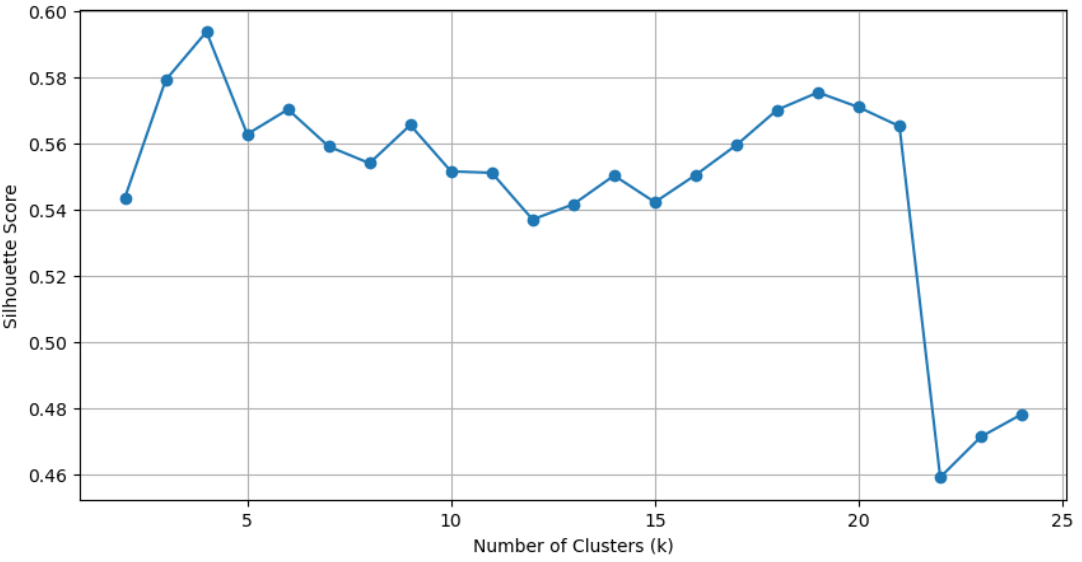}
        \label{fig:subfig2}
    }
    \hspace{0.02\textwidth}
    \subfigure[Optimal Cluster Selection Using Davies-Bouldin Index]{
        \includegraphics[width=0.4\textwidth]{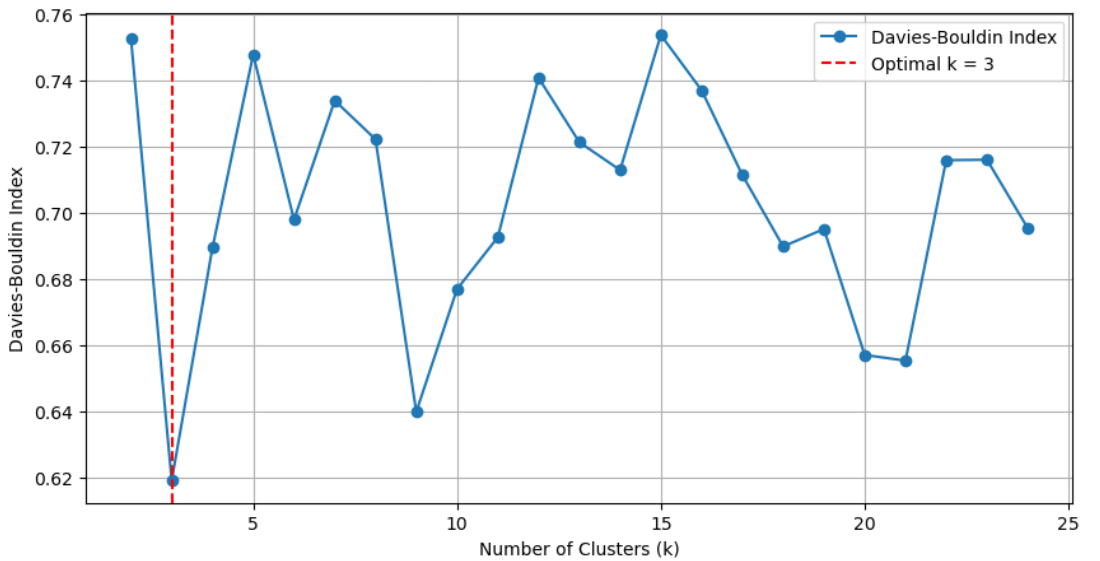}
        \label{fig:subfig3}
    }
    \caption{Evaluating optimal cluster counts for the input sensor data. Although the elbow method, silhouette, and Davies-Bouldin criteria yield different exact values ($k=6$, $k=4$, and $k=3$ respectively), their combined indication guides the model design.}
    \label{fig:mainfig}
\end{figure}

In order to understand the latent structure of the input sensor space and ensure an appropriate network topology, we applied $k$-means clustering to the original seven sensor features. Three standard metrics were used to find the optimal number of clusters, including: the elbow method, silhouette score, and Davies- Bouldin index~\cite{articlekmeans}. Figure~\ref{fig:subfig1} shows that the elbow method suggests $k=6$ due to the diminishing reduction in the within-cluster sum of squares, whereas the highest silhouette coefficient is reached at $k=4$ (Fig.~\ref{fig:subfig2}), and the Davies-Bouldin index is minimized at $k=3$ (Fig.~\ref{fig:subfig3}). 

\begin{figure}[H]
    \centering
    \subfigure[Variation of NRMSE]{
        \includegraphics[width=0.4\textwidth]{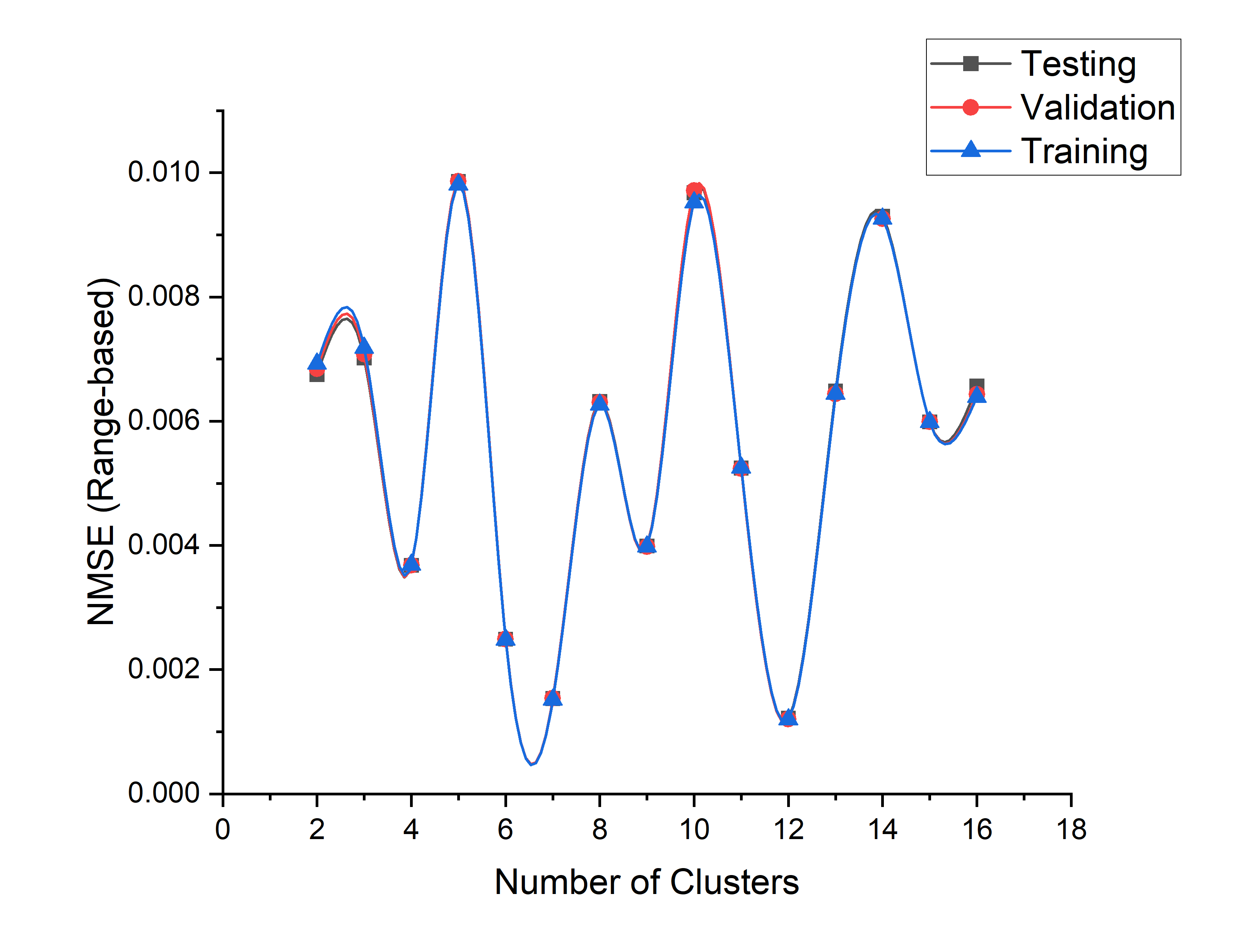}
        \label{fig:NRMSE}
    }
    \hspace{0.02\textwidth}
    \subfigure[Variation of $R^2$]{
        \includegraphics[width=0.4\textwidth]{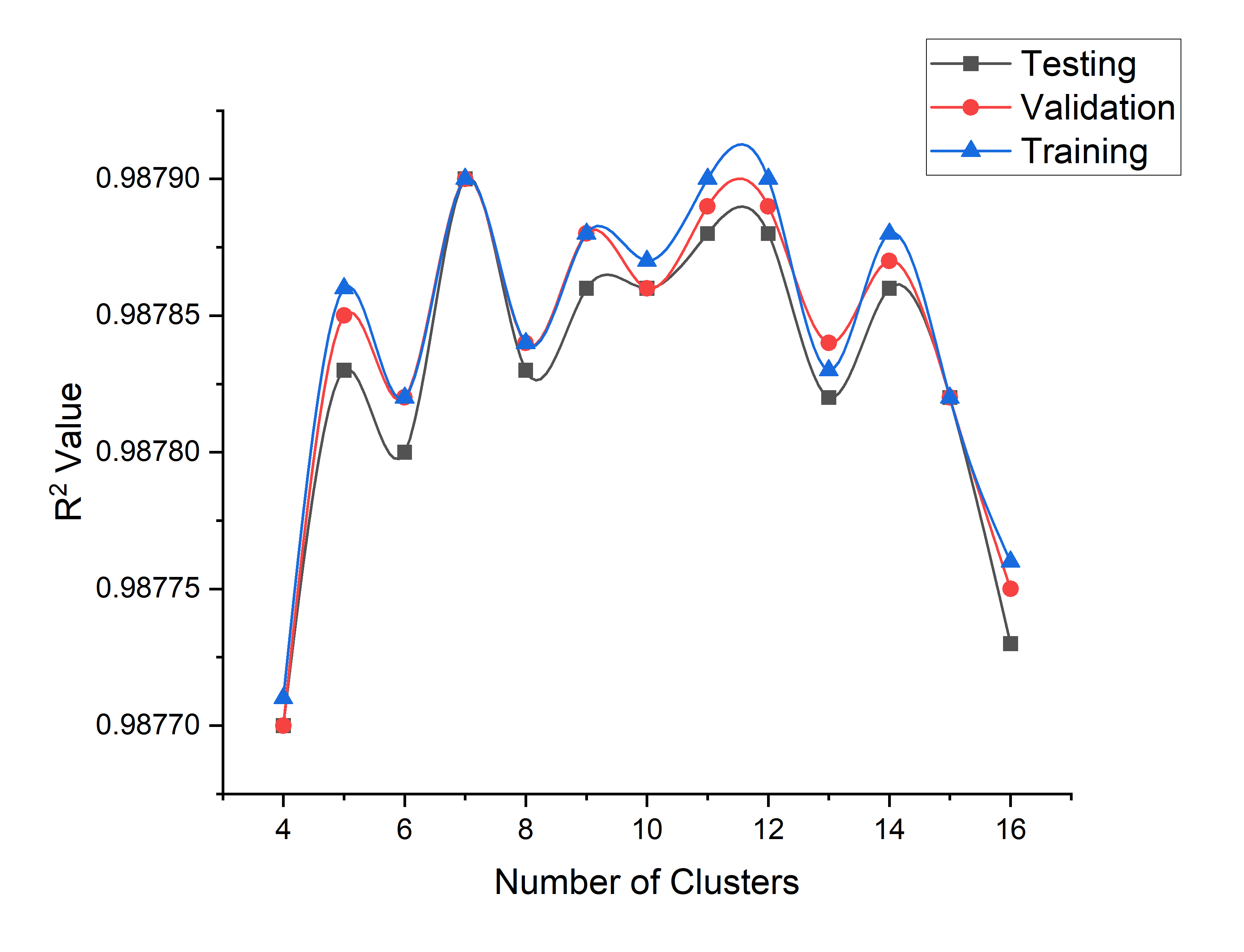}
        \label{fig:R2}
    }
    \caption{Influence of different cluster choices ($k$) on the MLP architecture. Using $k=7$ yields lower NRMSE and higher $R^2$, suggesting improved predictive accuracy and model alignment with data complexity.}
    \label{fig:clusterimpact}
\end{figure}

Given these mixed indicators, we next assessed the effect of varying $k$ on a \textit{classical} MLP’s performance in predicting displacements. Specifically, the number of clusters $k$ was used to set the neuron count in the final hidden layer, aiming to reflect the data’s natural segmentation. Figure~\ref{fig:NRMSE} and \ref{fig:R2} illustrate the normalized RMSE (NRMSE) and $R^2$ values for different $k$. Lower $k$ (e.g.\ 3 or 4) gave relatively lower $R^2$ and higher NRMSE. Meanwhile, too large a $k$ (above 7) appeared to cause over-segmentation, adding unnecessary complexity. Conclusively, $k=7$ achieved the best balance, improving both $R^2$ and NRMSE. Hence, we adopted 7 neurons in the last hidden layer of the optimized MLP used in subsequent experiments.

\subsection*{2. Quantum Feature Encoding and Comparative Model Performance}
\label{subsec:quantumcomparison}
After establishing an optimized classical MLP baseline, we explored multiple quantum-classical hybrid configurations. Table~\ref{tab:final_table} compiles their final performance metrics, including MSE, RMSE, $R^2$, and two measures of normalized RMSE (relative to data range and standard deviation). For clarity, we denote the following: \textbf{Baseline MLP} represents a purely classical network without quantum layers; \textbf{Quantum-Classical MLP} refers to our primary model where quantum layers are placed first, followed by two classical fully connected layers; \textbf{Classical-Quantum MLP} refers to the model with two classical fully connected layers first, then a parameterized quantum circuit, concluding with a final classical layer mapping an 8-qubit output to the 1017-dimensional displacement vector; \textbf{Classic MLP (Clustering enforced )} is extending Baseline MLP so clustering enforces the dimensionality of the mid layer of classic MLP; \textbf{QMLP (Poly-SPD) + clustering} refers to the version of thw proposed model, where polynomial feature expansion (degree 2 or 3) plus constructing a symmetric positive-definite (SPD) matrix for quantum embedding, when clustering enforces the dimensionality of the mid layer of classic MLP; \textbf{QMLP (Poly-SPD \& HC + clustering)} then extends the above with a Hilbert-Schmidt mapping to preserve geometric structure, effectively the \emph{most advanced} variant introduced.

\vspace*{1em}
\begin{table}[ht]
\centering
\caption{Comparison of Baseline MLP vs.\ Hybrid Quantum-Classical (QMLP) variants. All metrics were measured on the final test set.}
\label{tab:final_table}
\begin{tabular}{lccccc}
\toprule
\textbf{Model} & \textbf{MSE} & \textbf{RMSE} & \textbf{$R^2$} & \textbf{NRMSE(range)} & \textbf{NRMSE(std)} \\
\midrule
Baseline classic MLP
 & $1.008\times10^{-3}$ 
 & $3.1\times10^{-2}$ 
 & 0.9831
 & 0.00789 
 & 0.03513 \\

Quantum-Classical MLP
 & $4.72 \times 10^{-4}$
 & $2.217\times10^{-2}$
 & 0.9844
 & 0.01097
 & 0.04168 \\

Classical-Quantum MLP
 & $9.68 \times 10^{-4}$
 & $3.14\times10^{-2}$
 & 0.9614
 & 0.01104
 & 0.04394 \\

 Classic MLP (Clustering enforced )
 & $7.40\times10^{-10}$ 
 & $2.72\times10^{-5}$ 
 & 0.9878 
 & 0.00789 
 & 0.03513 \\

QMLP (Poly-SPD + clustering)
 & $4.65 \times 10^{-5}$
 & $6.82\times10^{-3}$
 & 0.9869
 & 0.00657
 & 0.02793 \\

\textbf{QMLP (Poly-SPD + HC + clustering)}
 & $\mathbf{3.16\times10^{-11}}$
 & $\mathbf{5.62\times10^{-6}}$
 & 0.9856
 & 0.00164
 & 0.00726 \\
\bottomrule
\end{tabular}
\end{table}
\vspace*{1em}

\paragraph{Baseline vs.\ Hybrid Variants }
Table~\ref{tab:final_table} compares all models on the final test set. The \emph{Baseline classic MLP} yields $R^2\approx0.9831$, while adding quantum layers (Quantum-Classical or Classical-Quantum MLP) shifts errors differently. Quantum-Classical obtains $4.72\times10^{-4}$ MSE, improving slightly over the baseline’s MSE exponent but showing a distinct RMSE drop. Classical-Quantum lags at $9.68\times10^{-4}$ MSE, implying that in this dataset, quantum-first strategies may outperform classical-front processing. Still, both hybrids stay competitive with purely classical designs.

\paragraph{Classic MLP with Clustering.}
Imposing cluster-based neuron counts notably enhances the classic MLP, cutting MSE down to $7.40\times10^{-10}$. This result ($R^2=0.9878$) surpasses the baseline’s $0.9831$, signifying that k-means–guided architecture tuning can yield significant gains even without quantum layers. This underscores how data segmentation helps represent varied operational regimes in structural response more effectively.

\paragraph{Value of Poly-SPD Embedding }
The QMLP (Poly-SPD + clustering) further reduces MSE to $4.65\times10^{-5}$ by coupling polynomial feature expansion with an SPD manifold prior to quantum encoding. Here, domain-aware expansions capture the system’s nonlinearities, and SPD conditioning stabilizes amplitude encoding. Consequently, the network achieves $R^2=0.9869$, matching or slightly exceeding classical baselines while remaining robust to sensor noise.

\paragraph{Best Performance: Poly-SPD + HC + Clustering.}
The strongest model, \emph{QMLP (Poly-SPD + HC + clustering)}, attains MSE $\approx 3.16\times 10^{-11}$ and RMSE of $5.62\times10^{-6}$. Although its $R^2$ (0.9856) is near but not strictly above the clustering-enforced MLP (0.9878), the absolute errors remain drastically lower, indicating tighter overall predictions. By integrating polynomial expansions, SPD encoding, Hilbert–Schmidt mappings (HC), and cluster-informed architecture, the hybrid pipeline fully exploits geometric embeddings and quantum transformations.

{\subsection{Computational Complexity}}

To quantify the extra work introduced by our quantum layers, we compare the per-sample inference cost of the baseline MLP with that of the hybrid QMLP.  A fully-connected MLP with input dimension \(d_{\rm in}=7\), two hidden layers of sizes \(h_{1}=64\), \(h_{2}=32\), and output dimension \(d_{\rm out}=1017\) requires
\[
C_{\rm classical}
=O\bigl(d_{\rm in}h_{1} + h_{1}h_{2} + h_{2}d_{\rm out}\bigr)
=O(7\cdot64 + 64\cdot32 + 32\cdot1017)\approx3.5\times10^{4}
\]
multiply–accumulate operations per sample.

Our QMLP adds four components: 
(1) a quadratic feature expansion to dimension \(d'=\binom{d_{\rm in}+1}{2}=28\), 
(2) construction and spectral decomposition of the \(d'\times d'\) SPD matrix at cost \(O(d'^{3})\) \cite{GolubVanLoan2013}, 
(3) vectorization into a quantum register of \(n=\lceil\log_{2}(d'^{2})\rceil=10\) qubits followed by \(L=10\) variational layers costing \(O(L\,n)\) two-qubit gates \cite{NielsenChuang2010}, 
and (4) a classical readout of size \(h_{3}=64\), costing \(O(n\,h_{3} + h_{3}\,d_{\rm out})\).  Summing these yields
\[
C_{\rm QMLP}
=O\bigl(d'^{3} + L\,n + n\,h_{3} + h_{3}\,d_{\rm out}\bigr)
=O(28^{3} + 10\cdot10 + 10\cdot64 + 64\cdot1017)\approx8.8\times10^{4}.
\]
Thus, the quantum‐layer overhead 
\[
\Delta C
= C_{\rm QMLP} - C_{\rm classical}
=O\bigl(d'^{3} + L\,n + n\,h_{3}\bigr),
\]
corresponds to an operation‐count ratio 
\[
R = \frac{C_{\rm QMLP}}{C_{\rm classical}} \approx 2.5.
\]
To put this into numbers, when MLP takes 15 ms to execute end-to-end, our proposed method takes 37.5 ms.

{
\subsection*{3. Implications for Real-Time SHM}
The proposed hybrid method achieves a significant computational speed-up compared to conventional FE modeling. A single FE analysis step in ANSYS approximately takes 30 seconds to simulate the structural response of this structure while the trained conventional MLP model takes 15 milliseconds and the QMLP incorporated MLP models need 37.5 milliseconds to predict the full-field displacement of the entire structure from the sensor readings and these models are 2000 times faster than the conventional FE modeling. However, the traditional FE modeling is forward in nature as it generates the structural response under known boundary conditions and loads. In contrast, the proposed QMLP-based method estimates the global response fields of the structure through sparse sensor measurements. Such an instant inverse calculation is critical in the context of digital twins as it leads to real-time structural health assessment, anomaly detection, and automated decision-making while drastically reducing the computational cost of traditional structural analysis processes. 
}

{
Achieving errors at magnitudes of $10^{-6}$ to $10^{-5}$ in RMSE is highly valuable for near–real-time structural health monitoring. Although this numerical improvement may appear modest, its practical implications are significant, especially for large-scale civil infrastructure \cite{doi:10.1139/cjce-2021-0627}. In our case study involving over 300 structural nodes, this level of accuracy translates to sub-millimeter displacement resolution, which is critical for early detection of anomalies and deterioration. It also contributes to reducing false positive alerts, thereby supporting more reliable decision-making in asset management systems. While quantum layer overhead is nontrivial, advancing hardware may mitigate it. The synergy of clustering, polynomial expansions, and SPD-based quantum encoding represents a promising route for scaling SHM to more complex structures with minimal loss of accuracy.
}

\section*{Limitations and Future Outlook }
{Though our demonstration addresses a medium‐scale structure, scaling to thousands of sensors will demand more advanced circuits (e.g.\ error mitigation on real hardware), temporal embeddings or recurrent variational layers for transient loads, and richer SPD constructions (e.g.\ kernel‐ or manifold‐learning). Implementing this framework on noisy devices remains challenging; nevertheless, SPD‐driven hybrid models show significant promise for robust, data‐driven SHM in the era of large‐scale digital twins. In future work, we will (i) compile device specs (qubit count, connectivity, gate fidelities, coherence times) to estimate Hilbert–Schmidt embedding resources; (ii) extend embeddings to time‐series via temporal kernels and recurrent circuits; (iii) explore alternative SPD maps (diffusion maps, graph kernels); (iv) conduct noise‐aware ablations under zero‐noise extrapolation and probabilistic error cancellation; (v) compare parameter efficiency and generalization of hybrid QMLP vs.\ classical MLP; (vi) implement key subroutines on 5–10 qubit hardware to benchmark fidelity and latency; and (vii) analyze asymptotic scaling to pinpoint qubit thresholds for quantum advantage.}

\bibliography{sample}

\section*{Author contributions statement}

A.A conceptualization of Quantum machine learning, Quantum machine learning experiment conducting, manuscript writing, analysis of the result, and supervision, S.J data generation, conceptualization, machine learning experiments conduction, analyzed the data, and wrote the manuscript. M.M., S.M, J.T and S.S reviewed and supervised the project.  

\section*{Additional information}

To include, in this order: \textbf{Accession codes} (where applicable); \textbf{Competing interests} (mandatory statement). 

The corresponding author is responsible for submitting a \href{http://www.nature.com/srep/policies/index.html#competing}{competing interests statement} on behalf of all authors of the paper. This statement must be included in the submitted article file.

\end{document}